**Distinct mechanisms of DNA sensing based on N-doped carbon nanotubes with enhanced conductance and chemical selectivity****

*Han Seul Kim, Seung Jin Lee, and Yong-Hoon Kim**

[*]     Prof. Y. -H. Kim, H. S. Kim
Graduate School of Energy, Environment, Water, and Sustainability, Korea Advanced Institute of Science and Technololgy
291 Daehak-ro, Yuseong-gu, Daejeon 305-701, Korea
E-mail: y.h.kim@kaist.ac.kr

        Prof. S. J. Lee
College of Pharmacy, Gachon University
Hambakmoeiro 191, Yeonsu-gu, Incheon 406-799, Korea

        Prof. Y. -H. Kim
KI for the NanoCentury, Korea Advanced Institute of Science and Technololgy
291 Daehak-ro, Yuseong-gu
Daejeon 305-701, Korea

Carrying out first-principles calculations, we study N-doped capped carbon nanotube (CNT) electrodes applied to DNA sequencing. While we obtain for the face-on nucleobase junction configurations a conventional conductance ordering where the largest signal results from guanine according to its high highest occupied molecular orbital (HOMO) level, we extract for the edge-on counterparts a distinct conductance ordering where the low-HOMO thymine provides the largest signal. The edge-on mode is shown to operate based on a novel molecular sensing mechanism that reflects the chemical connectivity between N-doped CNT caps that can act both as electron donors and electron acceptors and DNA functional groups that include the hyperconjugated thymine methyl group.

**1. Introduction**

Doping with foreign atoms represents the primary process of altering the properties of sp$^2$-bonded carbon network. In particular, nitrogen doping of graphene, graphene nanoribbons, and carbon nanotubes (CNTs) is an important modification process that has led to various potential device applications such as field-effect transistors,[1] fuel cells,[2] lithium ion batteries,[3] capacitors,[4] and molecular and biosensors.[5, 6] Significant progress in the fundamental scientific understanding of CNT or graphene N doping is currently being made as well.[7-9] Among the list of promising device applications based on carbon nanomaterials lies DNA sequencing.[10-16] However, there has been so far little work that considered applying N-doped CNTs or graphene to DNA sequencing.[16]

DNA sequencing, detecting and differentiating the four nucleobases in a DNA strand, adenine, cytosine, guanine, and thymine, is one of the most important goals in biology and has immense medical implications in that it can usher in a new era of personalized medicine. In this effort, the approach that has emerged as a promising new direction is the combination of the longitudinal translocation of single-stranded DNA (ssDNA) through solid-state nanopores and the simultaneous sensing of nucleobases using transverse electron tunneling.[17-20] After the initial theoretical propositions,[21, 22] the feasibility of tunneling-based identification of individual nucleotides[23] or nucleosides[24] has now been demonstrated through experiments. However, in these proof-of-principles experiments that employed Au or functionalized Au electrodes, signals from different nucleotides markedly overlapped with each other and the difficulty arises from the overadsorption of nucleotides to metallic surfaces, which makes it desirable to explore other possibilities such as graphene or CNT electrodes.

In this work, using a first-principles computational approach that combines density functional theory (DFT) and matrix Green's function (MGF) calculations, we study the application of the N-doped capped CNT electrodes to the sensing of the four nucleotides [Figure 1(a)], deoxyadenosine 5'-monophosphate (dAMP), deoxycytidine 5'-monophosphate (dCMP), deoxyguanosine 5'-monophosphate (dGMP), and deoxythymidine 5'-monophosphate (dTMP). After showing that one can obtain significantly enhanced transmissions by adopting N-doped capped CNTs as nanoelectrodes, we will demonstrate that configuring them to adopt small and large gap sizes, which accommodate the face-on and edge-on DNA base configurations, respectively, results in two completely different conductance orderings with the largest signals originating robustly from dGMP and dTMP in the former and latter cases, respectively. We analyze the mechanisms of these two sequencing modes and show that the two distinct orderings come from the opposite limits of competition between energetic and chemical connectivities: In the small-gap/face-on electrode/nucleobase configuration that maximizes the π-π stacking, the highest occupied molecular orbitals (HOMO) location or nucleobase size dictates the conductance ordering as was found in previous studies. On the other hand, in the large-gap/edge-on electrode/nucleobase configuration, we find that the chemical connectivity between nucleobase functional groups and N-doped CNT caps becomes the dominant factor. Since almost all of the proposed and realized sequencing methods ranked the signal from purine nucleotides, dGMP or dAMP, as the largest,[11, 13, 15-17, 22] the latter conductance ordering that ranks dTMP as the source of the largest signal is especially noticeable. It will be shown that the novel nucleobase sensing mode is possible due to the ability of N-doped CNTs that can behave both as electron donors and acceptors and the electron-withdrawing nature of the hyperconjugated methyl group.

**2. Results and Discussion**

**2.1. Junction models based on nitrogen-doped capped carbon nanotube electrodes**

As nanoelectrodes, we employed N-doped capped metallic (5,5) CNTs of 6.8 Å diameter [Figure 1(b)], which is comparable to the nucleobase spacing of ~ 7 Å in a stertched ssDNA[see Supporting Figure S1(a)]. Because the π-π interactions between CNT caps and DNA bases can be maximized and accordingly DNA conformational fluctuations and translocation speed can be reduced, the capped CNTs should be listed as a promising nanoelectrode candidate for DNA sequencing.[10, 15] We have previously shown that capped semiconducting CNTs have intrinsically low (Schottky barrier-free) metallic contact resistances due to the cap localized states that are well connected with the CNT body states.[25] However, in the metallic (*n*, *n*) armchair CNTs, the capped localized states are energetically located far away from the Fermi level ($E_F$) (at around 0.8 eV above and 1.3 eV below $E_F$ in our case) and the states that can carry currents around $E_F$ are minimal, as can be observed in the local DOS (LDOS) spectra shown in Figure 1(c). The transmission data obtained for these capped CNTs mirror-symmetrically placed at 3.5 Å gap distance are shown together and confirm that the localized CNT cap states have negligible contributions in inducing transmission near $E_F$. Given the necessity to improve the charge transport capacity of capped CNT electrodes, we considered further the substitutional doping of CNT caps with a nitrogen atom. The effect of N doping in introducing CNT cap-originated current carrying states near $E_F$ was also theoretically pointed out.[26] In Figure 1(d), we show the effects of introducing a substitutional N atom into capped (5,5) CNT (*capN1* model). The LDOS spectra show the appearance of strong *n*-type defect states (0.09eV above $E_F$) caused by the N p$_z$ orbital,[7] in addition to cap-originated localized states downshifted (compared with their pristine counterparts) to 0.38 eV and 0.55 eV above $E_F$. The transmission curve obtained for these *capN1* CNTs that mirror-symmetrically placed at the gap distance of 3.5Å are shown together in Figure 1(d), and comparison with Figure 1(c) clearly illustrates the drastic enhancement of transmission around $E_F$.

Junction models were then constructed by introducing dAMP, dCMP, dGMP, and dTMP into a gap between two N-doped capped CNT electrodes. We comment that, although we explicitly included part of the backbone composed of a nonplanar sugar ring and a phosphate group [Figure 1(a)] to extract realistic conformations of nucleobases with respect to the CNT electrodes, the current signal is minimally afffected by the backbone [see Supporting Figures S1(b) and S1(c)]. We chose the DNA geometries such that nucleobases are located near the midpoint of the CNT-CNT gap, which we set as the coordinate origin. This correspdonds to the assumption that the backbone translocates along the $x$-axis direction while it is controllably shifted away from the midpoint between electrodes along the $y$-axis direction [See, e.g., Figure 2(a)]. Note that, thanks to the diameter of electrodes, the transverse current signal along the CNT-axial $z$-axis direction is primarily determined by the nucleobase located between CNTs and minimally affected by the neighboring nucleobases [see Supporting Figure S1(a)]. As the junction geometries and associated DNA configurations, we considered two predominantly different cases: One with a small electrode-electrode gap distance ($l_{FO} = 6 \sim 8$ Å) that induces the "face-on" configurations of DNA bases, and the other with a large distance ($l_{EO} = 12 \sim 14$ Å) that accommodates the "edge-on" configurations of bases.

It should be emphasized that the parameter space we explored is very large. For the electrodes, we considered three gap distances for each face-on and edge-on configurations, three N doping sites [Figure 1(b)], and several different combinations of N-doped CNT electrodes and their relative orientations. The data presented below will be based on the heterogeneous *capN1-capN2* combination with the N atoms in the two CNT caps arranged centrosymmetrically with the midpoint between electrodes as the inversion center. Data obtained for other electrode combinations will be presented in Supporting Information. Thermal fluctuation of DNA is another important variable that should be properly taken into account.[17-20, 22] For the nucleotide configurations, starting from the initial reference geometry in which the center of mass of each nucleobase is placed at the center of the gap, we first scanned the translocation-direction $x$ axis by up to ± 4 Å with the displacement step of 0.5 Å. After identifying the reference nucleotide geometries, we next generated additional configurations mimicking thermal fluctuations by translating the nucleotides by ± 0.25 Å and ± 0.5 Å and also rotating them around the $x$, $y$, and $z$ axes by ± 5°. The selection criteria of reference geometries are different for the edge-on and face-on cases and will be detailed below.

**2.2. Small-gap "face-on" sequencing mode**

We first discuss the "face-on" configuration case that is schematically shown in Figure 2(a). As representative data, we present below the results obtained for the *capN1-capN2* and $l_{FO}$ = 6.5 Å electrode configuration. The potential wells along the translocation-direction $x$ axis are relatively deep (larger than $k_B T \approx 0.026$ eV) and have different energy minimum points depending on the nucleobases, as shown in Figure 2(b). We thus selected the energy minimum points as the reference geometries and generated additional sampling configurations. We found that the energy curves are not modified with the N atom doping sites within the CNT caps (see Supporting Figure S2), which indicates that the interactions between DNA bases and CNT caps in the face-on configuration are governed by the π-π interactions and only insignificantly modified by the N dopants.

Figure 2(c) shows the zero-bias-limit (Landauer formula) conductance histograms obtained for the 25 configurations for each nucleobase [and 50 for dAMP, since it has two local energy minima points. See Figure 2(b)] given in the unit quantum of conductance, $G_0$ = $2e^2/h$ = 77.6 S. The skewed histograms were fitted by Fisher-Tippet-Gumbel distribution function,[27] $f(x) = \frac{1}{\beta}\exp\left(\frac{x-\mu}{\beta}\right)\exp\left[-\exp\left(\frac{x-\mu}{\beta}\right)\right]$, with the standard deviation $\sigma = \beta\pi/\sqrt{6}$, where $\mu$ is the location parameter and $\beta$ is the scale parameter. We find relatively large conductance values and their ordering of dGMP > dAMP > dTMP > dCMP. However, the histograms significantly overlap with each other [$\log(G/G_0)$ = –2.46 ± 0.23, –2.96 ± 0.50, –3.38 ± 0.34, and –3.77 ± 0.41 for dGMP, dAMP, dTMP, and dCMP, respectively], which is in agreement with experimental findings [23, 24] (See also Supporting Figure S3 for other electrode combination cases).

To elucidate the origin of the face-on case conductance ordering, we observed DOS projected onto nucleotides and CNT electrodes, as shown in Figure 3(a). The corresponding transmission spectra are shown in Figure 3(b). The projected DOS of CNTs (red dotted lines) show two prominent peaks at the energies right above $E_F$ and 0.5 eV above $E_F$. As previously discussed [Figure 1(d)], the former is introduced by N doping and the latter originates from the downshifted CNT cap localized states. Both CNT states strongly hybridize with DNA states (solid blue lines), which thus enables the magnification of molecular signals around $E_F$. In addition, from the projected DOS of nucleotides, we find that the molecular signals originate from DNA HOMO states (indicated by red downward arrows) identified at, on average, 0.80 eV, 1.04 eV, 1.16 eV, and 1.16 eV below $E_F$ for dGMP, dCMP, dAMP, and dTMP, respectively. This ordering implies that the standard ordering of isolated DNA base HOMOs (dGMP > dAMP > dCMP > dTMP) was approximately maintained even after the coupling with electrodes. We emphasize that the HOMO ordering was a typical mechanism of previously realized nucleobase identification[23] and also supramolecular switches.[28, 29] However, we also note that in our case the HOMO ordering (dGMP > dCMP > dAMP > dTMP), with the exception of dGMP, does not directly translate into the conductance ordering (dGMP > dAMP > dTMP > dCMP). This inconsistency can be understood by reminding that the conductance determined by HOMO for the weak coupling limit is approximately given by [17, 30]

$$G \approx G_0 \times \frac{\gamma_1 \gamma_2}{(E_F - E_{HOMO})^2}, \qquad (1)$$

where $\gamma_{1/2}$ is the coupling strength (inverse lifetime) between a nucleotide and CNT electrode 1/2, which corresponds to the simplified scalar version of the broadening matrix $\Gamma_{1/2}$ that will be defined in Eq. (2), and observing that the degree of HOMO broadening that determines $\gamma_{1/2}$ is larger for purine nucleotides, dGMP and dAMP. That is, in addition to the dominant effects of energy level alignment, the physical size of nucleobases and their chemical coupling with CNT electrodes play a secondary role in the face-on configuration.

**2.3. Large-gap "edge-on" sequencing mode**

A practical solution to the limitation imposed by thermal fluctuations (statistical noises) and achieve reliable identification of all four nucleobases would be to introduce a second set of electrodes.[17, 19, 20, 22] Particularly, based on the above analysis of correlations between conductance and HOMO orderings that indicated a competition between energetic and chemical connectivities [$(E_F - E_{HOMO})^2$ and $\gamma_1 \gamma_2$ terms of Eq. (1), respectively] and the dominance of the former in the face-on mode, we explored a novel approach that can maximize the role of the latter chemical-connectivity mechanism. We now demonstrate that such a complementary method can be realized with the same N-doped CNT electrodes by probing the dominantly edge-on DNA base configurations at the increased gap size to $l_{EO}$ = 12.0 Å ~ 14.0 Å.

The *capN1-capN2* $l_{EO}$ = 14.0 Å device with the "edge-on" nucelobase configurations (that should produce the largest current signals for the large-gap electrode setting) is schematically shown in Figure 4(a). To generate junction models, we followed a procedure similar to that in the face-on case. In scanning the potential energy surfaces along the translocation trajectories, the main differences originating from employing increased gap distances are, as shown in Figure 4(b), much shallow potential energy curves. Given the negligible restrictions on the nucleotide configurations, we took the nucleotide-electrode configurations that correspond to the highest-transmission values indicated by arrows in Figure 4(c) as reference geometries [Figure 4(d)].

Transmission histograms generated with 25 edge-on junction models with $l_{EO}$ = 12.0 Å for each nucleotide case are shown in Figure 5(a). We emphasize that, since we considered only the edge-on nucleobase configurations in this large-gap $l_{EO}$ case, the histograms presented here should be regarded as the distributions of the maximum transmission values (see Supporting Figure S4). First, we note the conductance ordering of dTMP > dGMP ≈ dCMP > dAMP [$\log(G/G_0)$ = − 4.44 ± 0.27, − 5.61 ± 0.40, − 5.79 ± 0.35, and − 6.44 ± 0.38 for dTMP, dGMP, dCMP, and dAMP, respectively] is different from that of the face-on case and, to our knowledge, any other experimentally or theoretically reported orderings. The signals again significantly overlap with each other, but the resolution is much improved compared with the face-on counterpart. Figure 5(b) shows that this ordering is robustly maintained even for the increased gap distance of $l_{EO}$ = 14.0 Å [$\log(G/G_0)$ = − 6.11 ± 0.06, − 6.31 ± 0.05, − 6.34 ± 0.03, − 6.44 ± 0.03, and for dTMP, dGMP, dCMP, and dAMP, respectively]. Considering other electrode combinations (see Supporting Figure S5), we found that the conductance ordering of dTMP being the first and dAMP being the fourth is consistently maintained when *capN1* and *capN2* electrodes are involved, but it does not hold any more with the employment of *capN3*, i.e. the proximity between N doping atom and nucleobases are the key for this sequencing mode. Experimentally, it might be difficult to precisely place N dopant atoms near the end of CNT caps (*capN1* and *capN2*), but we anticipate that such CNT electrodes can be screened by, e.g., calibration runs using short deoxythymidine homopolymers.

Finally, we discuss the microscopic origin of the novel edge-on DNA sensing mechanism. As shown in Figure 6(a), the HOMO ordering in the edge-on case is that of the isolated nucleobases, dGMP > dAMP > dCMP > dTMP, which suggests that the sensing mechanism of the edge-on case is intrinsically different from that of the face-on case. The critical role of the edge-on nucleobase conformations and having the N dopant atom located near the end of CNT cap shows that the edge-on sensing mechanism is based on the chemical connectivity between N-doped CNT caps and nucleobases. To explicitly show that the nature of charge conduction paths is affected by the strength of chemical coupling $\gamma_1$ and $\gamma_2$ in the edge-on case, we show in Figure 6(b) the spatial distributions of LDOS near $E_F$ of the most conducting dTMP and the least conducting dAMP for the $l_{EO}$ = 14.0 Å case. In agreement with their conductance ordering, although the LDOS of dAMP is only connected with the left CNT through the amine group, the LDOS of dTMP is connected both with the left and right electrodes through the methyl and carbonyl groups, respectively. That is, the three functional groups present in nucleobases, amine, carbonyl, and methyl groups, establish an electron conduction path upon bridging the N dopant atoms embedded in CNT caps.

More detailed information on the chemical connectivity between the three nucleobase functional groups and N-doped CNT caps can be obtained by observing the charge transfer between them. The charge redistributions $\Delta\rho = \rho_{CNT+DNA} - (\rho_{CNT} + \rho_{DNA})$ for the dAMP and dTMP cases are shown in Figure 6(c). Similar plots for all four bases at a different isovalue are presented in Supporting Figure S6. One can observe the charge transfer from the electron donating amine group to CNT in dAMP, dCMP, and dGMP, and that from the CNT to the electron withdrawing ketone group in dCMP, dGMP, and dTMP. Namely, the ability of N-doped CNT electrodes to act both as electron donors and electron acceptors is an important prerequisite of the edge-on nucleobase sensing. Note that both amine and ketone groups mentioned above participate in the base paring through hydrogen bonding. Interestingly, for dTMP, the hyperconjugated methyl group[31] that is not involved in base pairing also withdraws electrons from CNT in our edge-on junction configuration. The resulting symmetric charge transfer (compared to the asymmetric ones in the other three bases) and the almost linear charge flow path [see Figures 4(d) and 6(b)] explains the unusual finding of the highest conductance through dTMP in the edge-on configuration.

## 3. Conclusions

In summary, we showed that the substitutional N doping of CNT electrodes results in greatly enhanced transmission *as well as* chemical sensitivity and thus allows two complementary DNA sequencing modes. Analyzing the sensing mechanisms, we found that the conductance ordering from the small-gap face-on electrode set, in which the dGMP signal is the largest, is mainly determined by the typical mechanism of nucleobase HOMO level positions. On the other hand, we have shown that the hitherto unreported conductance ordering obtained for the large-gap edge-on electrode set, where the dTMP signal is the largest, originates from the chemical connectivity between nucleobase functional groups and N-doped CNT caps. The latter novel edge-on sensing mechanism in which the methyl group of thymine (in addition to ketone and amine groups) plays an important role suggests its potential in probing the DNA methylation or nucleobase modifications in general, which has important implications for epigenetics research concerning gene expression, development and disease that includes cancer.[32, 33]

## 4. Experimental Section

*Computational methods*

All the calculations carried out in this work are fully first-principles based on the DFT and DFT-MGF formalisms. The DFT calculations were performed within Perdew-Burke-Ernzherof parameterization of the generalized gradient approximation[34] using a modified version of SeqQuest program[35] that employs norm-conserving pseudopotentials[36] and corresponding double-ζ-plus-polarized Gaussian basis sets. Charge transport characteristics of semi-infinite junctions were computed based on the DFT-MGF approach[30] as implemented in our in-house code:[28, 37] We first calculated Hamiltonian $H$ and overlap $S$ matrices for a device model and corresponding two separate infinite CNT electrode models, from which we extracted surface Green's functions. We then calculated the retarded Green's function, $G(E) = (ES - H + \Sigma_1 + \Sigma_2)^{-1}$, where the self-energies, $\Sigma_{1/2}$, describe in an *ab initio* fashion energy level shift and broadening of nucleotides interacting with the CNT electrodes 1/2. Transmission functions were calculated according to

$$T(E) = Tr[\Gamma_1(E)G(E)\Gamma_2(E)G^+(E)], \qquad (2)$$

where $\Gamma_{1/2} = i(\Sigma_{1/2} - \Sigma_{1/2}^+)$ are the broadening matrices. The corresponding electronic structures were analyzed using the charge density differences and projected and local density of states (DOS) plots.

**Acknowledgements**
    We thank Dr. Sang Joo Lee (Asan Medical Center) and Prof. Sang Bok Lee (University of Maryland) for valuable discussions. This research was supported by EDISON Program (No. 2012M3C1A6035684), Basic Science Research Grant (No. 2012R1A1A2044793), Nano·Material Technology Development Program (2012M3A7B4049888), and 2013-Global Ph.D. Fellowship Program of the National Research Foundation funded by the Ministry of Education, Science and Technology of Korea. Computational resources were provided by the KISTI Supercomputing Center (KSC-2012-C2-20).

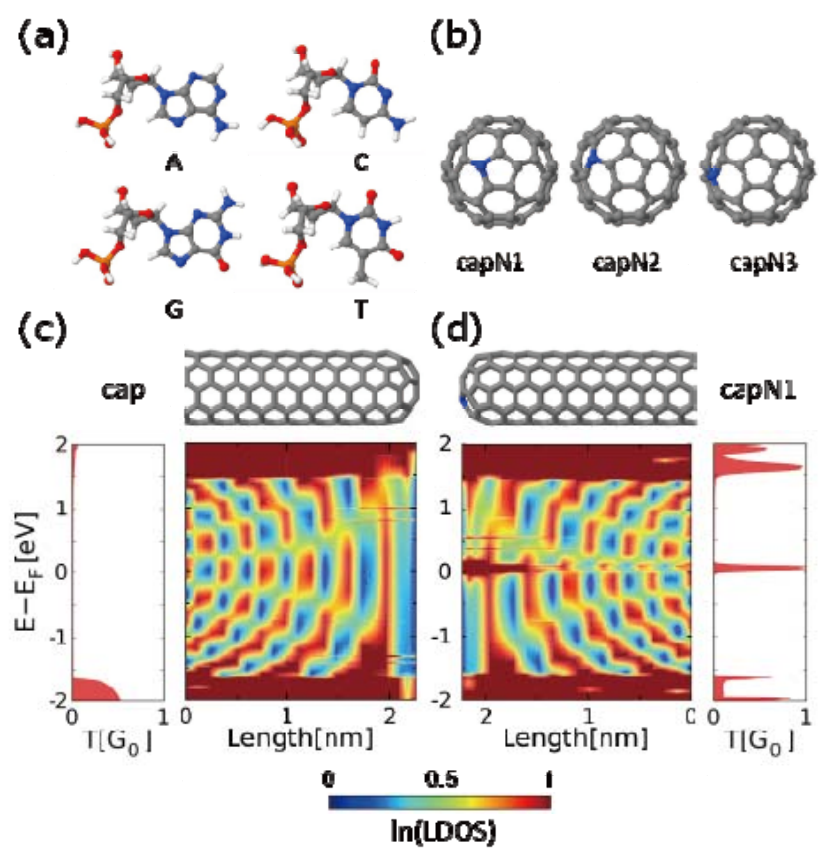

**Figure 1.** Atomic models of DNA nucleotides and N-doped capped CNT electrodes with enhanced conductance and chemical sensitivity. (a) The four DNA nucleotide models: dAMP, dCMP, dGMP, and dTMP. (b) N-doped capped CNTs with the N dopant atom located at the first (*capN1*), second (*capN2*), and third (*capN3*) carbon rings from the cap end. Transmission and LDOS of the (c) pristine and (d) N-doped capped CNT electrodes.

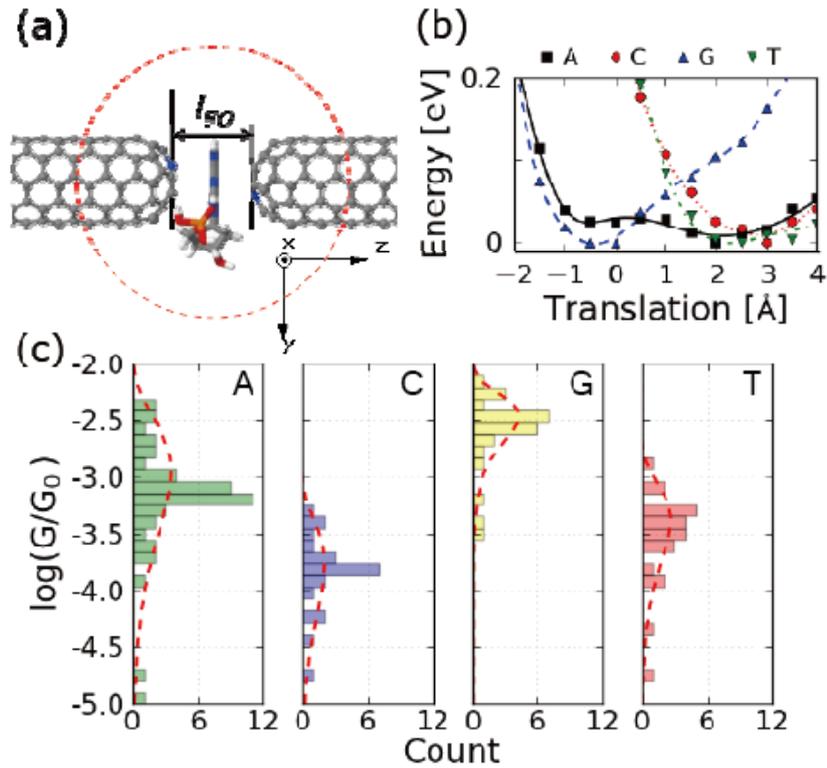

**Figure 2.** (a) A schematic device configuration of the "face-on" DNA sequencing mode, in which two *capN1* CNT electrodes with the gap size $l_{FO}$ = 6.5 Å is attached (or embedded) to a nanopore (indicated by a dashed red circle) through which a ssDNA is translocated. (b) Potential energy surfaces of the four nucleobases sandwiched between *capN1-capN2* CNT electrodes along the translocation x-axis direction. (c) The zero-bias-limit conductance histograms corresponding to the thermally fluctuating nucleotides at their respective energy-minimum points shown in (b).

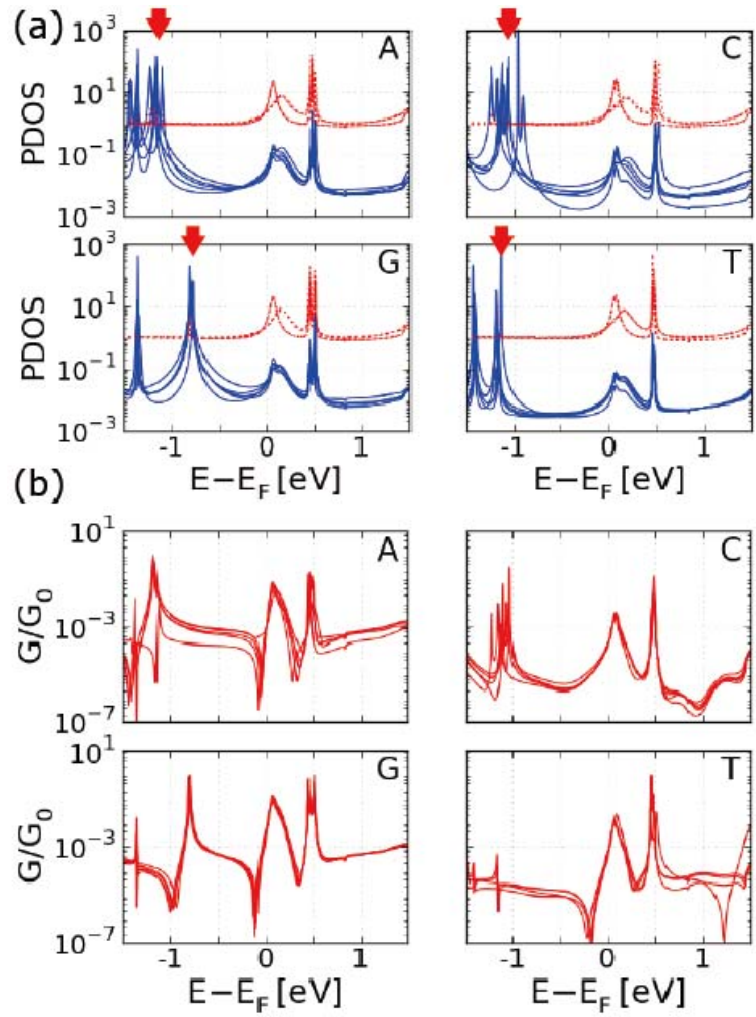

**Figure 3.** Composites of the (a) projected DOS of CNTs (dotted red lines) and nucleotides (solid blue lines) and (b) transmission functions corresponding to five representative face-on *capN1-capN2* junction configurations at $l_{FO}$ = 6.5 Å. Red arrows in (a) denote the approximate HOMO level locations in the four nucleotide cases.

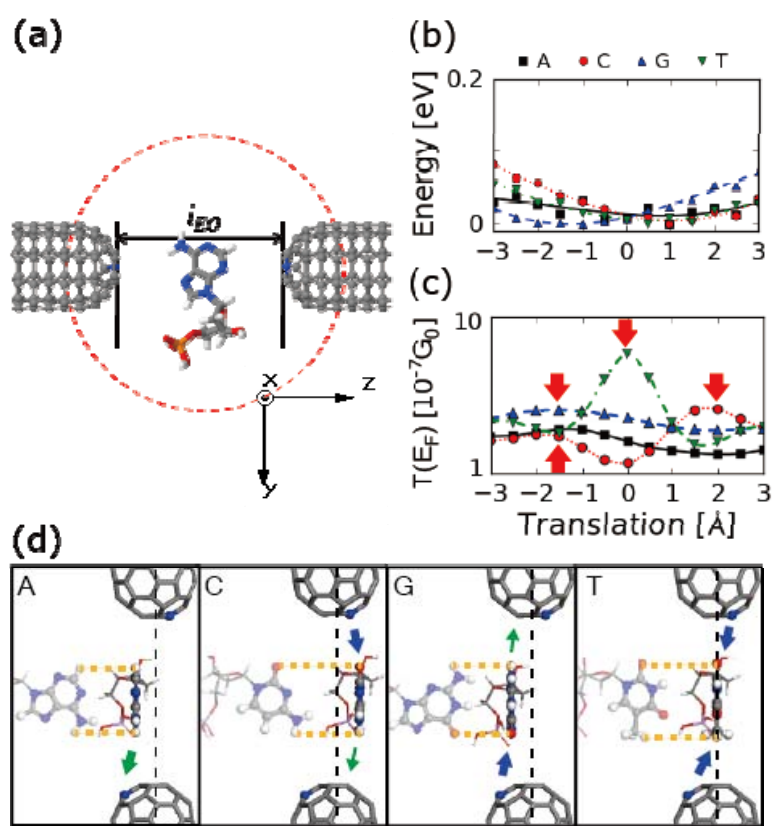

**Figure 4.** (a) A schematic device configuration of the "edge-on" DNA sequencing mode. The dashed red circle indicates a nanopore through which a ssDNA is translocated. (b) Potential energy surfaces and (c) zero-bias-limit conductances of the four nucleotides translocating along the x-axis direction. (d) Atomic structures of the junction models at the nucleotide positions where the maximum transmission values are obtained [denoted by red arrows in (c)]. Blue and green arrows indicate the CNT-to-DNA and DNA-to-CNT electron transfer directions, respectively. The thick and thin arrows indicate the major and minor couplings between the N-deoped CNT caps and the nucleobase functional groups, respectively.

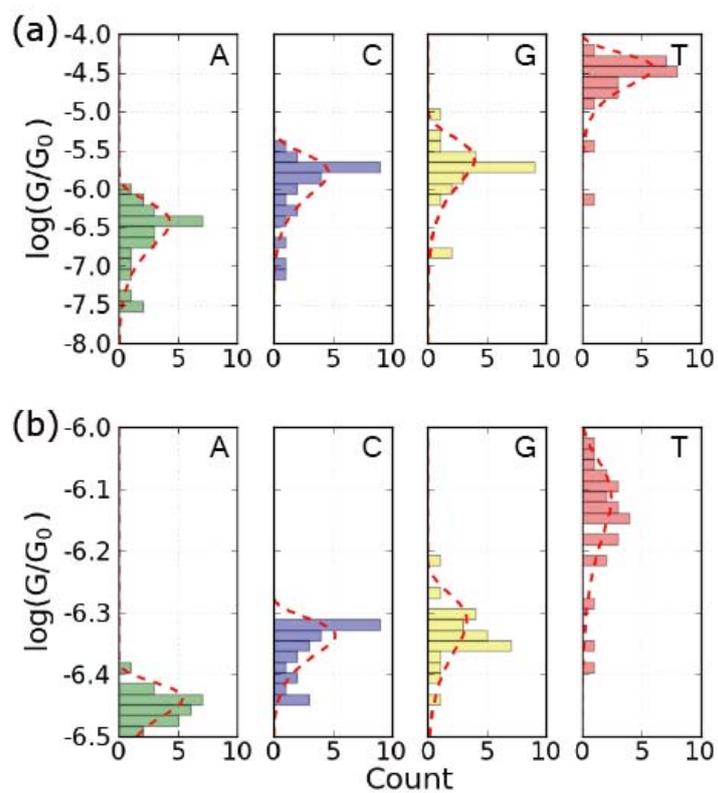

**Figure 5.** Zero-bias-limit conductance histograms of the four nucleotides for the edge-on *capN1-capN2* cases with (a) $l_{EO}$ = 12.0 Å and (b) $l_{EO}$ = 14.0 Å.

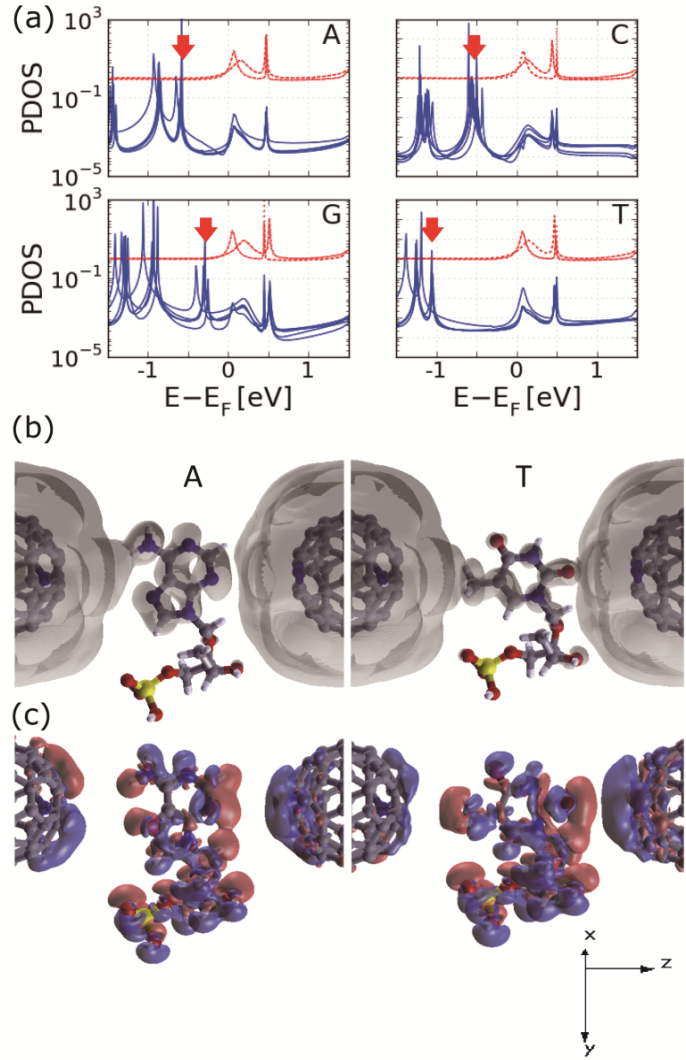

**Figure 6.** (a) Composites of the PDOS of CNTs (dotted red lines) and nucleotides (solid blue lines) for the $l_{EO}$ = 14.0 Å *capN1-capN2* edge-on cases. (b) Spatial distributions of the LDOS around $E_F$ for the $l_{EO}$ = 14.0 Å *capN1-capN2* edge-on dAMP and dTMP junctions at the isovalue of $1\times10^{-7}$ Å$^{-3}$eV$^{-1}$. (c) Corresponding charge density difference distributions at the isovalue of $5\times10^{-5}$ e·Å$^{-3}$. Red and blue colors represent positive (electron accumulation) and negative (electron depletion) isovalues, respectively.

**The table of contents entry: Nitrogen-doped carbon nanotube electrodes with the ability to act both as electron donors and electron acceptors allows a novel DNA sequencing mechanism based on the chemical connectivity between nitrogen doping sites and the functional groups of molecules.**

TOC Keyword: Carbon nanotubes, Doping, DNA recognition, Biosensors, *Ab initio* calculations

H. S. Kim, S. J. Lee, Y. -H. Kim*

Title
Distinct mechanisms of DNA sensing based on N-doped carbon nanotubes with enhanced conductance and chemical selectivity

ToC figure

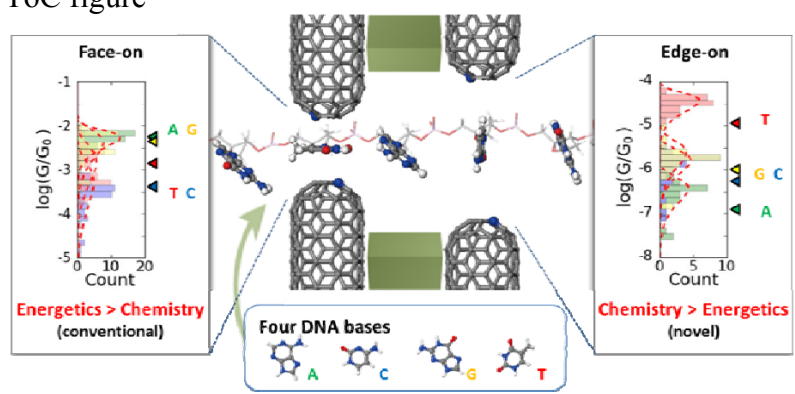

Page Headings
Left page:   H. S. Kim et al.
Right page:   DNA sensing based on N-doped carbon nanotubes

**Supporting Information**

# Distinct mechanisms of DNA sensing based on N-doped carbon nanotubes with enhanced conductance and chemical selectivity

Han Seul Kim,[1] Seung Jin Lee,[2] and Yong-Hoon Kim[1,3,*]

*1 Graduate School of EEWS, KAIST, 291 Daehak-ro, Yuseong-gu, Daejeon 305-701, Korea*

*2 Asan Medical Center, University of Ulsan College of Medicine, 388-1 Pungnap-dong, Songpa-gu, Seoul 138-736, Korea*

*3 KI for the NanoCentury, KAIST, 291 Daehak-ro, Yuseong-gu, Daejeon 305-701, Korea*

**Supporting Methods**

To calculate the geometries and corresponding electronic structures of DNA molecules sandwiched between N-doped capped (5,5) carbon nanotube (CNT) electrodes, we performed density functional theory (DFT) calculations using a modified version of SeqQuest program (Sandia National Labs.).[1] We used the Perdew-Burke-Ernzerhof (PBE) parameterization of the generalized gradient approximation[2] as the exchange-correlation functional. For pseudopotentials, we employed Hammann-Schlüter-Chiang (for C, H and P)[3] and Troullier-Martins (for O and N)[4] norm-conserving semilocal pseudopotentials. We adopted the double $\zeta$-plus-polarization (DZP)-level Gaussian basis sets for the DNA molecules and cap region of CNT electrodes, while single-$\zeta$ (SZ)-level basis sets for the bulk part of CNT electrodes. We confirmed that replacing the DZP basis sets with the SZ basis sets for the bulk CNT regions does not affect the computational results. Real-space integration was done with the grid spacing of 0.3 Bohr and only a single Γ $\vec{k}$-point was sampled.

To compute the electron transport properties of the CNT-DNA-CNT junction models, we carried out DFT-based matrix Green's function (MGF)[5] calculations using our in-house code.[6, 7] Transmission function was calculated according to

$$T(E) = Tr[\Gamma_1(E)G(E)\Gamma_2(E)G^+(E)]$$ (1)

where $G(E)$ is the retarded Green's function,

$$G(E) = (ES - H + \Sigma_1 + \Sigma_2)^{-1}$$ (2)

and Γ is the broadening matrix,

$$\Gamma_{1,2} = i(\Sigma_{1,2} - \Sigma_{1,2}^+).$$ (3)

Here, $S$ and $H$ are respectively the overlap and Hamiltonian matrices corresponding to the channel region (molecular part). The self-energy matrices,

$$\Sigma_{1,2} = x_{1,2} g_{S1,2} x_{1,2}^+,$$ (4)

where $x_{1,2}$ are the molecule-electrode 1 or 2 contact parts of the total $ES - H$ matrices and $g_S$ are surface Green's functions, rigorously describe the shift and broadening of molecular energy levels due to the coupling with electrodes. The surface Green's function $g_S$, the key quantity necessary to calculate the self-energy accurately, were extracted from two independent DFT calculations for the two unit-cell (5,5) bulk CNTs corresponding to the electrode 1 and electrode 2 regions using a 100 $\vec{k}_\perp$-point sampling along the CNT axial direction. In calculating $T(E)$, the energy was scanned from 2.0 eV below to 2.0 eV above the Fermi level with the 0.01 eV step. The nature of electronic structures and conductance channels were analysed through projected density of states (PDOS) and spatial distributions of local density of states and charge density differences.

**Supporting Figures**

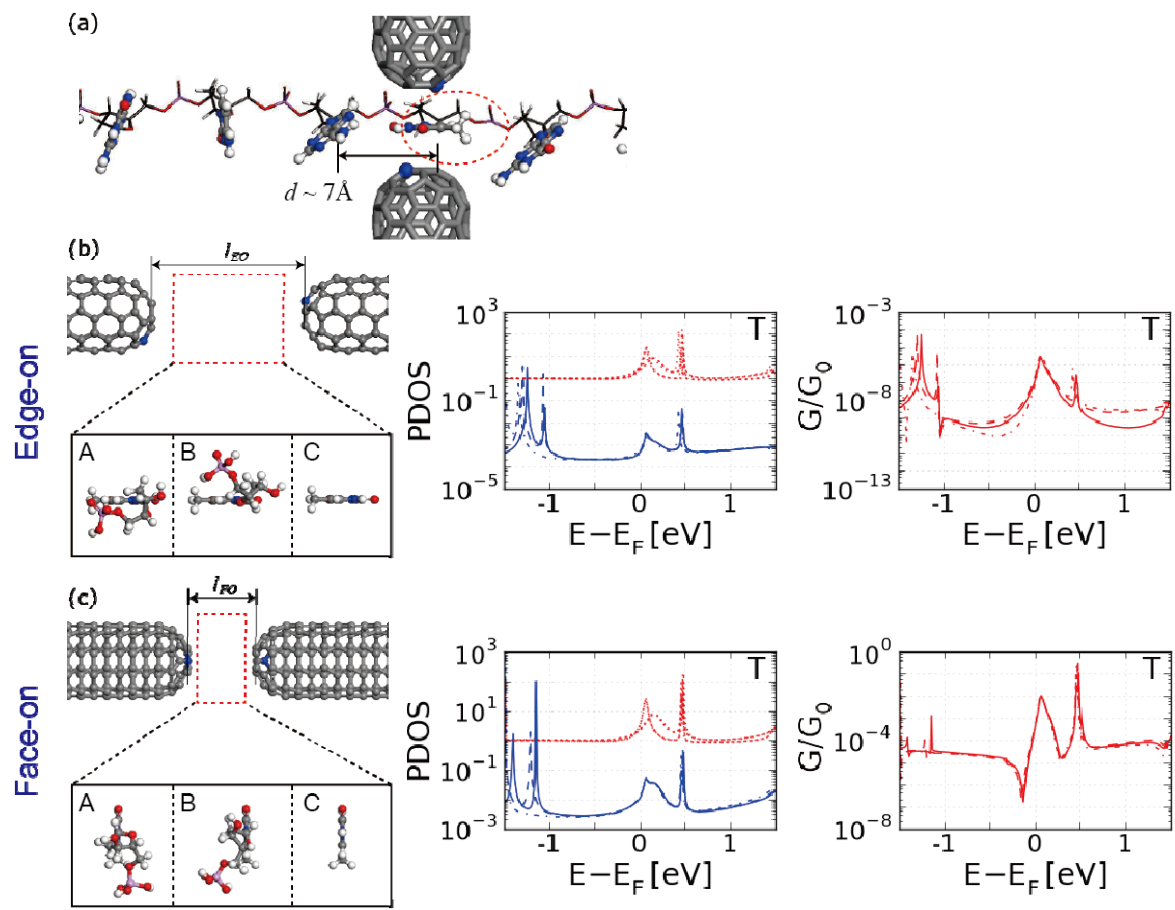

**Figure S1.** (a) A schematic figure that depicts the situation in which a ssDNA translocates through a nanopore and subsequently a CNT-CNT nanogap attached to the nanopore. The dotted elipse denotes a nucleotide section that is explicitly included in the first-principles quantum mechanical calculations. The spacing between nearest-neighbor nucleobases was estimated as ~ 7 Å. Various T DNA models that are different in terms of the presence/absence and conformation of the backbone (A: a dTMP extracted from a double-stranded B-DNA, B: a dTMP in which the backbone was twisted from A by about 45°, and C: a T nucleobase-only model) were placed in the (b) $l_{EO}$ = 14.0 Å "edge-on" and (c) $l_{FO}$ = 6.5 Å "face-on" *capN1-capN2* electrode gaps. In both cases, projected DOS of CNTs (red lines) and DNA bases (blue lines), and transmission functions computed with A, B, and C models (solid, dashed, and dot-dashed lines, respectively) are shown together.

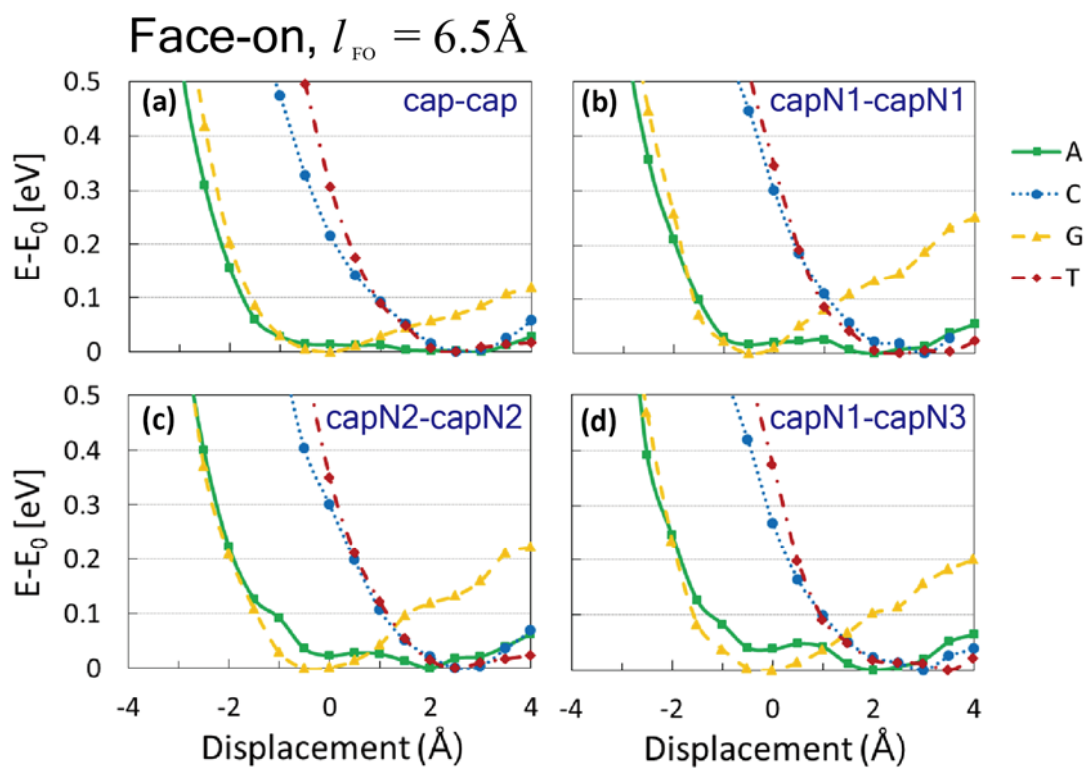

**Figure S2.** Potential energy curves obtained from the four nucleotides that translocate through the small-gap $l_{FO}$ = 6.5 Å CNT-CNT gaps at four different *capN* electrode combinations.

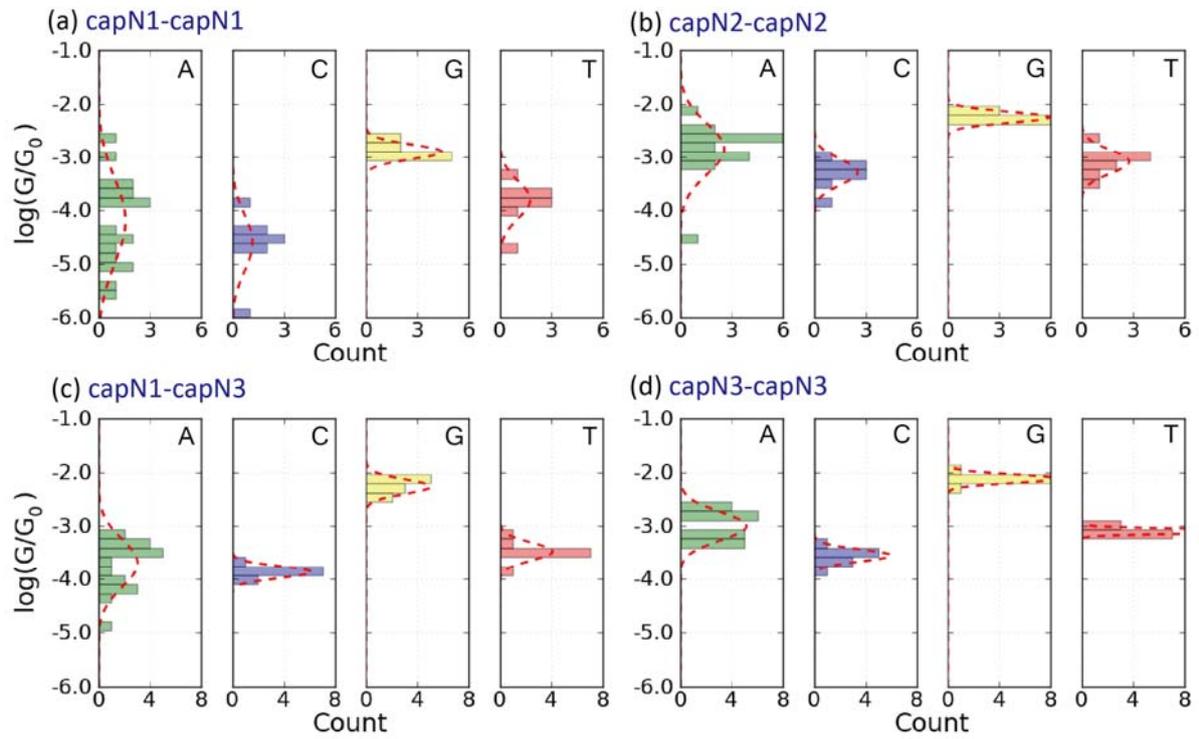

**Figure S3.** Conductance histograms for the face-on case with $l_{FO}$ = 6.5 Å in four different electrode combinations. Each histogram was constructed from 10 ~ 20 samples.

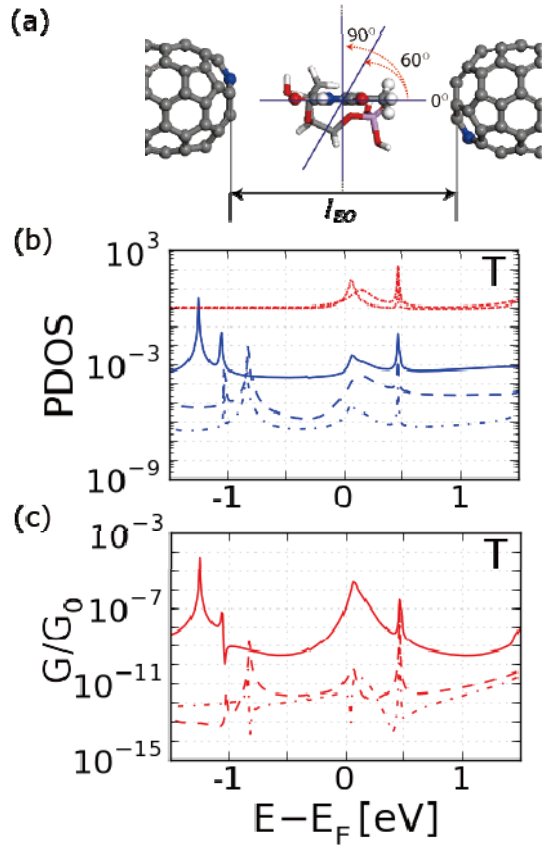

**Figure S4.** (a) A large-gap $l_{EO}$ = 14.0 Å *capN1-capN2* electrode device model, with which the nucleobase orientatoin effects were tested by rotating a dTMP from the "edge-on" 0° to "face-on" 90° nucelobase configurations. Resulting (b) projected DOS of CNTs (red lines) and DNA bases (blue lines) and (c) transmission functions. Solid, dashed, and dot-dashed lines respectively denote the 0°, 60°, and 90° cases.

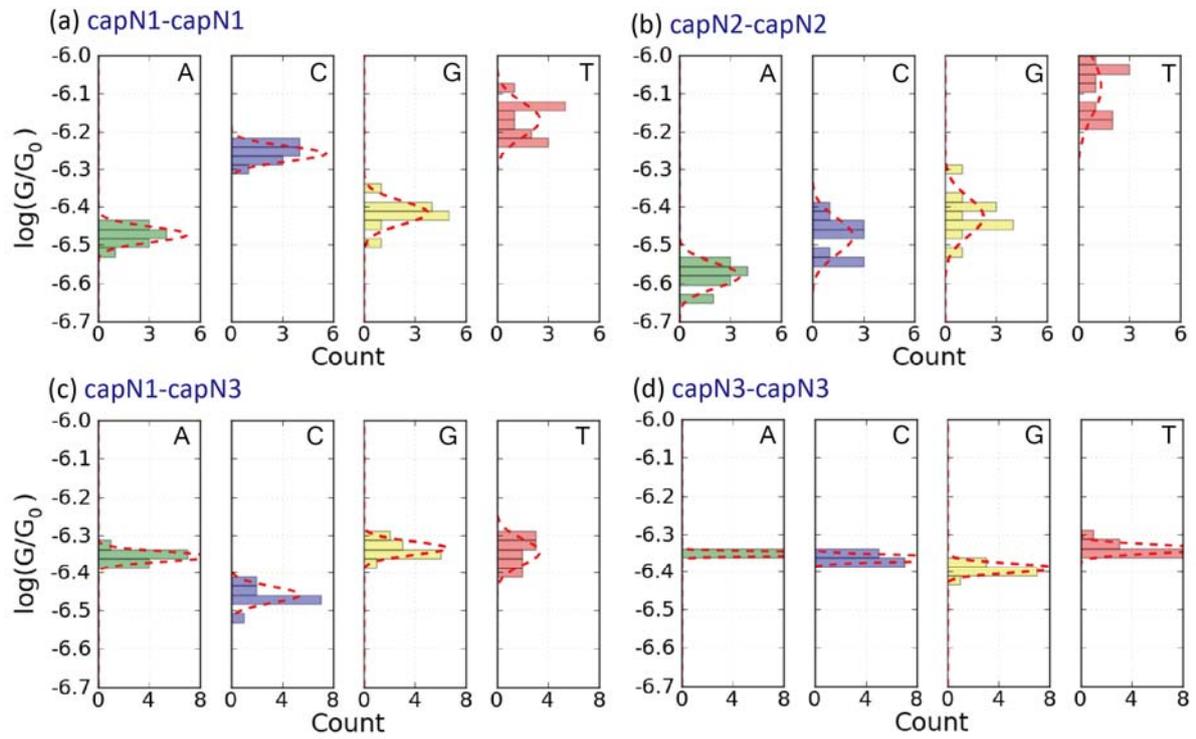

**Figure S5.** Conductance histograms for the edge-on case with $l_{EO}$ = 14.0 Å in four different electrode combinations. Each histogram was constructed from 13 samples.

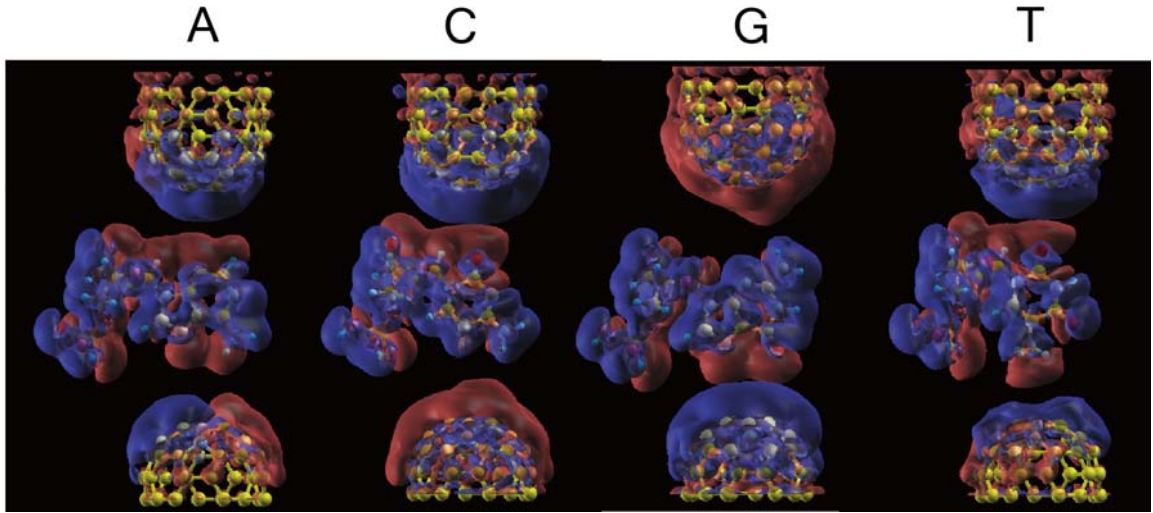

**Figure S6.** Spatial distributions of charge density differences between the edge-on junction model and the corresponding isolated DNA and CNTs, $\Delta\rho = \rho_{CNT+DNA} - (\rho_{CNT} + \rho_{DNA})$, for the isovalue of $2 \times 10^{-5}$ e·Å$^{-3}$. Each configuration corresponds to the maximum-conductance case. Red and blue colors represent the positive (electron accumulation) and negative (electron depletion) isovalues, respectively.